\documentclass[letter]{jpsj2} 
\title{Quasiparticle Interactions for f$^2$-Impurity Anderson Model with Crystalline-Electric-Field : Numerical Renormalization Group Study}

\author{Kazumasa \textsc{Hattori}$^{1}$\thanks{E-mail address: hattori@blade.mp.es.osaka-u.ac.jp}, Satoshi \textsc{Yotsuhashi}$^{2}$ and Kazumasa \textsc{Miyake}$^{1}$}

\inst{$^{1}$Division of Materials Physics, Department of Materials Engineering Science, Graduate School of Engineering Science, Osaka University, Toyonaka, Osaka 560-8531, Japan \\
$^{2}$Correlated Electron Research Center, 
the National Institute of Advanced Industrial Science and Technology, 
Tsukuba 305-8562 and Advanced Technology Research Laboratories, 
Matsushita Electric Industrial Co., Ltd., Kyoto 619-0237}

\abst{The aspect of the quasiparticle interaction of a local Fermi liquid, the impurity version of f$^2$-based heavy fermions, is studied by the Wilson numerical renormalization group method. In particular, the case of the f$^2$-singlet crystalline-electric-field ground state is investigated assuming the case of UPt$_3$ with the hexagonal symmetry. It is found that the interorbital interaction becomes larger than the intraorbital one in contrast to the case of the bare Coulomb interaction for the parameters relevant to UPt$_3$. This result offers us a basis to construct a microscopic theory of the superconductivity of UPt$_3$ where the interorbital interactions are expected to play important roles.}

\kword{f$^2$, numerical renormalization group, impurity Anderson model, CEF singlet ground state, Fermi liquid, Non-Fermi liquid}

\begin{document}
\maketitle

Heavy fermions containing two f-electrons at each site (f$^2$) have attracted
 much attention for a decade or so. Although uranium compounds such as URu$_2$Si$_2$, UBe$_{13}$, and UPt$_3$, and Pr-based filled skutterudites exhibit rich phenomena such as superconductivity, anomalous magnetism, non-Fermi liquid (NFL), metal-insulator transition and so on, a microscopic theory for these systems does not seem to have been well developed. Coming to think of low temperature physics of those f$^2$ systems, particularly superconductivity, we need to know  on the basis of a microscopic approach how quasiparticles of f$^2$-based compounds interact with each other. In this paper, we employ the numerical renormalization group (NRG) method \cite{Wilson,Krishna2} at the level of impurity problems considering the case of UPt$_3$\cite{UPt3Rev} with the hexagonal symmetry, because the local part of quasiparticle interaction is expected to play a crucial role also in the lattice problem. Our basic assumption is that the ground state of a crystalline-electric-field (CEF) is the f$^2$-singlet. The Pt Knight shift experiment \cite{tou,tou2} and the magnetic succestibility of UPd$_2$Al$_3$ \cite{grauel}, which has the same crystal structure as UPt$_3$, indicate this assumption is reasonable.

In this paper, we use a two orbital Anderson model as an appropriate one for a system with the f$^2$-singlet ground state under the hexagonal symmetry. The validity of this choice is discussed in detail in the following.


The f$^1$-states split into three doublets in the $J$=$5/2$ manifold. When we assume a strong spin-orbit interaction, we can neglect the $J$=$7/2$ states and use a j-j coupling scheme to obtain the f$^2$-states. On the other hand, if the Hund-rule coupling is stronger then spin-orbit coupling, the f$^2$-states are constructed in Russell-Saunders coupling scheme. The states with $J$=$4$ split into five different symmetry states.

\begin{figure}[!t]
  \begin{center}
    \includegraphics[width=.5\textwidth]{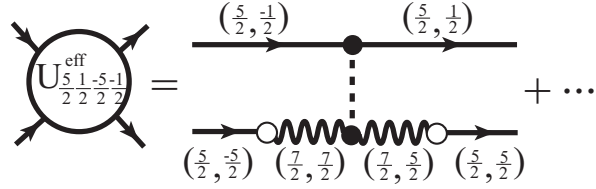}
  \end{center}
  	\caption{ Simplest diagram that generates the interaction of eq. (\ref{RenoInt}). $(J,J_z)$ denote the angular momentum $J$ and its z-component $J_z$. The straight lines represent the $J=5/2$ states and the wavy lines correspond to the $J=7/2$ states.The dashed line and open dots represent the Coulomb interaction and the CEF term, respectively.}
  	\label{fig:hyb}
  \end{figure}

 We assume the $\Gamma_4$ singlet ($[|3\rangle -|-3\rangle ]/\sqrt{2}$) is a ground state as mentioned above \cite{grauel}. The problem is how we connect these Hilbert spaces, which are in the opposite limit. When we construct the f$^2$-states in the j-j coupling scheme using only $J$=$5/2$ states, we obtain no singlets, because the states in the $J$=$5/2$ manifold under the hexagonal symmetry are the eigen states of the z-component of the total angular momentum operator $J_z^{\rm tot}$ (matrix element of $O_6^6$ is zero for $J=5/2$ states\cite{Stevens, Hutch}). Thus, the direct j-j coupling scheme using only the $J$=$5/2$ manifold is not appropriate in this case. In this paper, we use the $J$=$7/2$ states as intermediate states effectively to obtain f$^2$-singlet states. This stems from the fact that in the $J$=$7/2$ manifold, the matrix element of $O_6^6$ is not zero. This fact generates the interaction term, which do not conserve $J_z^{\rm tot}$ in the $J$=$5/2$ manifold as we eliminate the $J$=$7/2$ states from the consideration of Hilbert space(see Fig. \ref{fig:hyb}). This enables us to obtain a singlet ground state and we write it as
\begin{equation}
   U_{\frac{5}{2}\frac{1}{2}\frac{-5}{2}\frac{-1}{2}}^{\rm eff}
   f_{\frac{5}{2}}^{\dagger}
   f_{\frac{1}{2}}^{\dagger}f_{\frac{-5}{2}}f_{\frac{-1}{2}}+\rm h.c. \ \ .\label{RenoInt}
\end{equation}
In eq. (\ref{RenoInt}), we ignore retardation and $f_{j_z}$ is the destruction operator of the f-electron in the $J$=$5/2$ manifold. The magnitude of $U_{\frac{5}{2}\frac{1}{2}\frac{-5}{2}\frac{-1}{2}}^{\rm eff}$ is expected to be small. A detailed analysis of this term will be published elsewhere.
Other interactions such as the Coulomb interaction between the f-electrons and the CEF mixing are renormalized. However, their symmetry could not be changed. Note that eq. (\ref{RenoInt}) is the only allowed new interaction under the hexagonal symmetry in the effective $J=5/2$ manifold.
  \begin{table}[t]
    \begin{center}
  	\begin{tabular}{|l|l|}
  	\hline
  	$f^1 \otimes f^1$ & $f^2$\\
  	\hline
  	$\Gamma_7 \otimes \Gamma_9 $ & $ \Gamma_5 \oplus \Gamma_6$\\
  	$\Gamma_7 \otimes \Gamma_8 $ & $ \Gamma_3 \oplus \Gamma_4\oplus \Gamma_5$\\
  	$\Gamma_8 \otimes \Gamma_9 $ & $ \Gamma_5 \oplus \Gamma_6$\\
  	\hline
  	\end{tabular}
  	\end{center}
  	\caption{Irreducible decomposition of direct products under $D_{3h}$. Only the different orbital combinations are listed. $j_z=\pm5/2 (\Gamma_8), \ j_z=\pm1/2 (\Gamma_7), \ j_z=\pm3/2 (\Gamma_9)$}
  	\label{tbl:Grouop}
  \end{table}

Next, we restrict ourselves to the $J_z$=$\pm5/2,\pm1/2$ states of $J$=$5/2$ manifold, because $J_z=\pm3/2$ states have no contributions to the f$^2$-singlet states(see Table \ref{tbl:Grouop}). Other states are not important in the discussion of the low-energy quasiparticle properties under our assumption; f$^2$-singlet ground state. Indeed, many band calculations suggest that the $J=5/2$ states are dominant around Fermi surfaces for UPt$_3$\cite{Harima,Harima2,Zwicknagl}.

 Before writing our Hamiltonian, we transform representations from $j_z$ to pseudo-spin. This is performed using 
  \begin{figure}[t]
  \begin{center}
    \includegraphics[width=.2\textwidth]{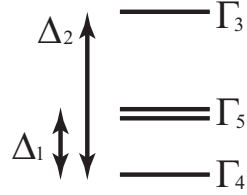}
  \end{center}
  	\caption{f$^2$ level scheme. $\Delta_1$ and $\Delta_2$ are related to $K_z$ and $K_{\pm}$ in eq. (\ref{Hund}) as $K_z = \Delta_2-2\Delta_1$ and $K_{\pm} = -\Delta_2$, respectively.}
  	\label{fig:level}
  \end{figure}
\begin{eqnarray}
  f_{\frac{5}{2}} &\to& f_{A\uparrow}, \ \ \ 
f_{-\frac{5}{2}} \to f_{A\downarrow},\\
f_{-\frac{1}{2}} &\to& f_{B\uparrow},\ \ \ 
f_{\frac{1}{2}} \to f_{B\downarrow},
  \end{eqnarray}
where $f_{j_z}$ and $f_{\alpha\mu}$ are f-electron field operators with the z-component of angular momentum $j_z$, and $\alpha$ and $\mu$ are the orbital and pseudo-spin indices, respectively.
In this representation, the f$^2$-atomic energy spectra are given by an anisotropic antiferromagnetic Hund's-rule coupling\cite{Yotsu}(See Fig. \ref{fig:level}). The splitting of $\Delta_2$ originats from eq. (\ref{RenoInt}) (redefined as $U_{\frac{5}{2}\frac{1}{2}\frac{-5}{2}\frac{-1}{2}}^{\rm eff} \equiv K_{\pm}/2$). 

Then our Hamiltonian is
\begin{eqnarray}
H &=&  H_{\rm c}+ H_{\rm loc}+ H_{\rm hyb}+H_{\rm Hund},%
\label{2orbIAM}\\
     H_{\rm c}   &=& \sum_{k\alpha\mu} %
                      \epsilon_k c_{k\alpha\mu}^{\dagger}c_{k\alpha\mu}, \label{cond}\\
     H_{\rm loc} &=& \sum_{\mu\alpha}\epsilon_{f_{\alpha}}%
                          f_{\alpha\mu}^{\dagger}f_{\alpha\mu}%
                    +\sum_{\alpha} U_{\alpha}n_{f_{\alpha\uparrow}}%
                                             n_{f_{\alpha\downarrow}}\nonumber\\
                    &&+U_{AB}\sum_{\mu\nu}n_{f_{A\mu}}n_{f_{B\nu}},\\
     H_{\rm hyb} &=& \sum_{k\mu\alpha}%
                      [V_{k\alpha} c_{k\alpha\mu}f_{\alpha\mu}+ {\rm h.c.}],\label{hyb}\\
     H_{\rm Hund} &=& -K_z S_A^z \cdot S_B^z - \frac{K_{\pm}}{2}[S_A^+ S_B^-%
                                      +S_A^- S_B^+]\label{Hund},
\end{eqnarray}
where $S_{\alpha}^z,S_{\alpha}^{\pm}$ are the z- and transverse components of the f-electron pseudospin operators of the $\alpha$(=A,B) orbital. The other notations are conventional. In eq. (\ref{hyb}) we have assumed that each f-electron mixes with the conduction electrons with the same point group symmetry and two conduction electron bands (from $-D$ to $D$) do not mix with each other. Note that the exchange interactions among different f-orbitals at the impurity site are not important for discussing the f$^2$-singlet ground-state model, because f$^2$-states include such terms in their energy levels. The Hamiltonian (eq. (\ref{2orbIAM})) can be regarded as a ``simplified" two-impurity Anderson model, although it contains an anisotropic spin-spin interaction and two nonequivalent impurities. The word ``simplified" means that the conduction electrons belonging to one band do not mix to those of the other band after impurity scattering.

In order to solve the Hamiltonian eq. (\ref{2orbIAM}) by the NRG method, we discretize conduction electron bands logarithmically and transform eq. (\ref{cond}) into one-dimensional semi-infinite chains, following Wilson \cite{Wilson}. Then we write eqs. (\ref{cond}) and (\ref{hyb}) as
\begin{eqnarray}
{H}_{\rm c} &=&\sum_{\alpha\mu}%
                       \sum_{n=0}^{\infty}\Lambda^{-n/2}t_{n}^{\alpha}[%
         f_{n\alpha\mu}^{\dagger}f_{n+1\alpha\mu}+%
       {\rm h.c.}]\label{2orbNRGconH},\\
     {H}_{\rm hyb} &=& \sum_{\alpha\mu}v_{\alpha}%
       \big[f_{-1\alpha\mu}^{\dagger}f_{0\alpha\mu}+%
       f_{0\alpha\mu}^{\dagger}f_{-1\alpha\mu}\big],\\
    t_n^{\alpha} &=& \frac{(1+\Lambda^{-1})(1-\Lambda^{-n-1})}%
          {2\sqrt{(1-\Lambda^{-2n-1})(1-\Lambda^{-2n-3})}},\\
v^2_{\alpha}&=& \sum_k |V_{k\alpha}|^2,
\end{eqnarray}
where $f_{-1\alpha\mu}\equiv f_{\alpha\mu}$, $f_{n\alpha\mu}\ (n \ge 0)$ is related to the conduction electron field operator via unitary transformations, and we take $D$ as the unit of energy. In this paper, we take the logarithmic discretization parameter as $\Lambda=2.5$.

 The Hamiltonian eq. (\ref{2orbIAM}) describes the competition between the Kondo-Yosida singlet and the f$^2$-CEF singlet. There are two stable Fermi liquid (FL) fixed points: the Kondo-Yosida-singlet fixed point and the CEF-singlet one. Between these two FL fixed points, there exists an unstable non-Fermi liquid (NFL) fixed point \cite{Yotsu}. Around the above two stable FL fixed points, we can write the effective Hamiltonian as
\begin{eqnarray}
H_{\rm eff}^N&=& H^{\ast}+\sum_{\alpha=A,B}\big[ \tilde{t}_{\alpha}\hat{O}_{1\alpha}%
                                    + \tilde{U}_{\alpha}\hat{O}_{2\alpha}\big]%
               + \tilde{U}_{AB}\hat{O}_3 \nonumber\\
               &&+ \tilde{K}_z\hat{O}_4 + \tilde{K}_{\pm}\hat{O}_5, %
               \label{2orbeffH}
\end{eqnarray}
where $\tilde{t}_{\alpha}, \tilde{U}_{\alpha}, \tilde{U}_{AB}, \tilde{K}_z$ and $\tilde{K}_{\pm}$ are parameters that should be determined by fitting the NRG spectrum near the FL fixed point whose Hamiltonian is $H^{\ast}$ (in this case, free Hamiltonian), $N$ means the number of conduction electron shells in the NRG procedure and $\hat{O}$'s are defined as
\begin{eqnarray}
\hat{O}_{1\alpha}&\equiv& \Lambda^{\frac{N-1}{2}}f_{0\alpha\mu}(u_{\alpha}f^{\dagger}_{0\alpha\mu}+t_{0}^{\alpha}f^{\dagger}_{1\alpha\mu})+ {\rm h.c.},\\
\hat{O}_{2\alpha}&\equiv& \Lambda^{\frac{N-1}{2}}(f^{\dagger}_{0\alpha\mu}f_{0\alpha\mu}-1)^2,\\
\hat{O}_3&\equiv& \Lambda^{\frac{N-1}{2}}(f_{0A\mu}^{\dagger}f_{0A\mu}-1)(f_{0B\nu}^{\dagger}f_{0B\nu}-1),\\
\hat{O}_4&\equiv& \frac{\Lambda^{\frac{N-1}{2}}}{4}(f_{0A\uparrow}^{\dagger}f_{0A\uparrow}-f_{0A\downarrow}^{\dagger}f_{0A\downarrow})\nonumber\\
&&\ \ \ \ \ \ \ \times(f_{0B\uparrow}^{\dagger}f_{0B\uparrow}-f_{0B\downarrow}^{\dagger}f_{0B\downarrow}),\\
\hat{O}_5&\equiv& \frac{\Lambda^{\frac{N-1}{2}}}{2}(f_{0A\uparrow}^{\dagger}f_{0A\downarrow}f_{0B\downarrow}^{\dagger}f_{0B\uparrow}\nonumber\\
&&\ \ \ \ \ \ \ \ +f_{0A\downarrow}^{\dagger}f_{0A\uparrow}f_{0B\uparrow}^{\dagger}f_{0B\downarrow}),
\end{eqnarray}
where $u_{\alpha}$ is the potential scattering amplitude of orbital $\alpha$ at the  FL fixed point and we use the Einstein contraction notation for $\mu$ and $\nu$.
This type of effective Hamiltonian was discussed in terms of the two-impurity Kondo problem except for the particle-hole and spin rotation symmetry \cite{JonesVerma}. Using eq. (\ref{2orbeffH}) we can extract quasiparticle interactions as
\begin{eqnarray}
F_{a}^{\alpha\alpha}&\equiv&z_{\alpha}^2%
        \Gamma_{\sigma\bar{\sigma}\sigma\bar{\sigma}}%
        ^{\alpha\alpha\alpha\alpha}=%
        \frac{(1+\Lambda^{-1})}{4}\frac{\tilde{U}_{\alpha}}%
        {\tilde{t}_{\alpha}^2},\label{QGam1}\\
  F_s^{AB}&\equiv&z_Az_B \sum_{\sigma} %
     \Gamma_{\uparrow\sigma\uparrow\sigma}^{ABAB}%
        =%
        \frac{1+\Lambda^{-1}}{4}\frac{\tilde{U}_{AB}}%
        {\tilde{t}_A\tilde{t}_B},\label{QGam2}\\
  F_a^{AB}&\equiv&z_Az_B \sum_{\sigma}%
       \sigma\Gamma_{\uparrow\sigma\uparrow\sigma}^{ABAB}%
        =%
        \frac{1+\Lambda^{-1}}{4}\frac{\tilde{K}_z/4}%
        {\tilde{t}_A\tilde{t}_B},\label{QGam3}\\
  F_{\rm ex}^{AB}&\equiv&z_Az_B%
        \Gamma_{\uparrow\downarrow\downarrow\uparrow}^{ABAB}=%
        \frac{1+\Lambda^{-1}}{4}\frac{\tilde{K}_{\pm}/2}%
        {\tilde{t}_A\tilde{t}_B}.\label{QGam4}
\end{eqnarray}
Here, $\Gamma_{\sigma_1\sigma_2\sigma_3\sigma_4}^{\alpha_1\alpha_2\alpha_3\alpha_4}$ is a full vertex part whose energy variables are set to zero (on the Fermi surface), $z_{\alpha}\propto \tilde{t}_{\alpha}^{-1}$ is a renormalization factor of orbital $\alpha$, and $\uparrow(\downarrow)$ corresponds to $\sigma=+1(-1)$. These results are obtained by comparing the form of pseudospin susceptibility and specific heat between NRG and local Fermi liquid theory. Note here that eq. (\ref{QGam4}) is not an exact relation, because the transverse pseudospin components are not conserved (we take the z-axis as the quantization axis), so that the contribution from incoherent parts remains finite. Although we cannot write $F_{\rm ex}$ only in terms of quasiparticles, in the NRG calculation, we think eq. (\ref{QGam4}) is a reasonable expression \cite{consider}.
Noted that the interaction $F_{\rm ex}^{AB}$ stems from the CEF and does not conserve the z-component of the total angular momentum.

  \begin{figure}[t!]
  \begin{center}
    \includegraphics[width=.45\textwidth]{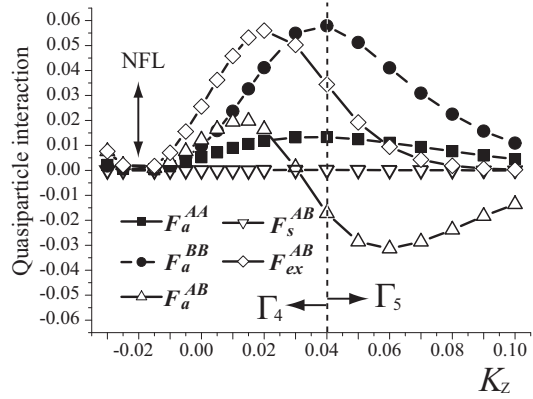}
  \end{center}
  	\caption{Quasiparticle interactions vs $K_z$. The parameters used in the calculation are $\epsilon_A=-0.6,\ \epsilon_B=-0.5,\ U_A=U_B=2.0,\ U_{AB}=0.4,\ v_A=v_B=0.25$ and $K_{\pm}=-0.04$. The NFL fixed point is $K_z^c\sim -0.02$ for these parameters. Note that $\Delta_1$ is zero at $K_z=-K_{\pm}=0.04$.}
  	\label{fig:Gam}
  \end{figure}
  \begin{figure}[t!]
  \begin{center}
    \includegraphics[width=.45\textwidth]{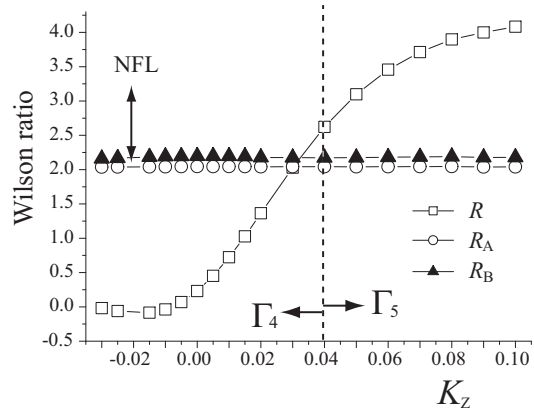}
  \end{center}
  	\caption{Wilson ratio vs $K_z$. $R_{\alpha}$ is the Wilson ratio of orbital $\alpha$ without interorbital contributions. $R$ is the total Wilson ratio. The parameters used in calculation are the same as in the Fig.\ref{fig:Gam}. These are calculated by peudo-spin representation, so that the results are different from {\it real} Wilson ratio.}
  	\label{fig:WilsonR}
  \end{figure}
The results of quasiparticle interactions and Wilson ratio are shown in Figs. \ref{fig:Gam} and \ref{fig:WilsonR}, respectively. Although we have performed calculations for various parameters, we have only shown the results for a specific parameter set because other sets also show qualitatively the same behavior in the Kondo regime.
Note that the level of excited $\Gamma_5$ coincides with that of $\Gamma_4$ at $K_z=-K_{\pm}$ and the level splitting between $\Gamma_5$ and $\Gamma_4$ increases as $K_z$ decreases. The interesting point in Fig. \ref{fig:Gam} is that the interorbital interactions $F_{a}^{AB}$ and $F_{\rm ex}^{AB}$ are enhanced around the intermediate coupling region ($K_z\sim 0.01$) and $F_{\rm ex}^{AB}$ becomes considerably larger than the intraorbital interactions. For a large $K_z$, this system approaches the fixed point of the anisotropic $S=1$ two-channel Kondo model, where the f-electron renormalization factors $z_{\alpha}\ (\alpha=A,B)$ become small. At the NFL fixed point ($K_z=-0.02$), $z_{\alpha}$ vanishes, which means the quasiparticle description breaks down.

 In Fig. \ref{fig:WilsonR}, both $R_A$ and $R_B$ are almost constant and close to 2 (see the caption of Fig. \ref{fig:WilsonR} for the definition of $R_{\alpha}$). This means intraorbital quasiparticle interaction is proportional to the renormalization factor ($F_{a}^{\alpha\alpha} \sim z_{\alpha} \sim T_{K_{\alpha}}$, where $T_{K_{\alpha}}$ is the Kondo temperature of orbital $\alpha$). This is a well-known in the two-impurity Kondo problem \cite{JonesVerma,AffLud}.

 There exist four parameters describing low-energy physics,i.e., $T_{K_{\alpha}}$ $(\alpha=A,B)$, $F_a^{AB}$ and $F^{AB}_{\rm ex}$. $F^{AB}_s$ is almost zero in the Kondo regime where the charge susceptibility and orbital susceptibility are suppressed.
  \begin{figure}[t]
  \begin{center}
    \includegraphics[width=.4\textwidth]{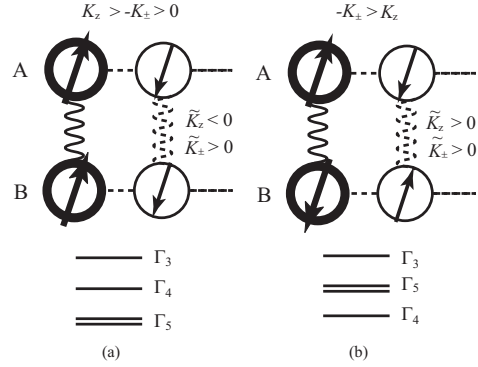}
  \end{center}
  	\caption{f- and conduction electron configuration for the case of (a) $\Gamma_5$ ground state and (b) $\Gamma_5$ ground state. The circles with thick and thin lines represent f-electrons and conduction electrons, respectively. The $\Gamma_5$ doublet state is constructed by $f_{A\uparrow}^{\dagger}f_{B\uparrow}^{\dagger}|0\rangle $ and  $f_{A\downarrow}^{\dagger}f_{B\downarrow}^{\dagger}|0\rangle$.
For the $\Gamma_4$ singlet, $\frac{1}{\sqrt{2}}[f_{A\uparrow}^{\dagger}f_{B\downarrow}^{\dagger} -f_{A\downarrow}^{\dagger}f_{B\uparrow}^{\dagger}]|0\rangle$.}
  	\label{fig:diss}
  \end{figure}

A tendency of interorbital couplings, $F_{a(\rm ex)}^{AB}$, reflects the property of the ground-state configuration of f-electrons (f$^2$- $\Gamma_4$, $\Gamma_5$). If the f$^2$-CEF ground state is the $\Gamma_5$ doublet (see Fig. \ref{fig:diss}(a)) and the antiferromagnetic coupling between the f-electrons and conduction electrons (Kondo type exchange) is taken into account, we obtain effective ferromagnetic coupling between the different orbital quasiparticles, consistent with $F_{a}^{AB}<0$ in Fig. \ref{fig:Gam}. The same argument as that in the case of the $\Gamma_5$ ground state is possible for the $\Gamma_4$-singlet ground-state case, see Fig. \ref{fig:diss}(b). However, the region where the $\Gamma_5$-doublet is the ground state, is not valid in the present model because of Hund's rule \cite{Caut}. In the case of f$^{2}$-$\Gamma_4$-singlet ground states, we need not worry about Hund's rule, because our f$^2$-singlet states have only $J=4$ components. In addition, due to the reduction of $z_{\alpha}$ toward the NFL fixed point the interorbital couplings $F_{{a({\rm ex})}}^{AB}$ are suppressed as shown in Fig. \ref{fig:Gam}. Around the intermediate region, the interorbital quasiparticle interactions ($F_{a(\rm ex)}^{AB}$) are well enhanced compared with the intraorbital ones $(F_{a}^{AA},\ F_{a}^{BB})$, which decrease monotonically toward the NFL fixed point. It is interesting that although the overall energy scale becomes small as we approach the NFL fixed point, which is a direct consequence of $z_{\alpha}\to0$, the details of interactions i.e. anisotropies vary.

 The above results imply that the interactions between quasiparticles, in a system with multiorbitals (specifically f$^2$-singlet ground state), are quite different from a simple model, such as a simple-minded multiorbital Hubbard model, in general. If we use the renormalized interactions obtained here to tackle the f$^2$ heavy fermion superconductors, there exists a possibility of obtaining spin triplet superconductivity because of large interorbital interactions\cite{Takimoto}. 
Several theories of triplet superconductivity induced by Hund's rule have been proposed thus far \cite{Norman,hotta}. Our results are, however, different from those in the point that the relevant interactions are those for quasiparticles (in impurity version) and their interactions originate from the low-energy CEF level f-electron. Although the contributions of $J_z=\pm 3/2$ have not been included in the present model, we believe they would not change the quantitative features of our results. The actual calculation of the superconducting transition temperature of UPt$_3$ is very complicated, because of two inequivalent sites in the unit cell and the orbital degrees of freedom per site. This will be discussed elsewhere.

 In summary, we have discussed the quasiparticle interactions of a local Fermi liquid with the f$^2$-singlet ground state under the hexagonal symmetry. We have found by the NRG method that interorbital interactions are much larger than intraorbital ones in the wide range of the CEF. This result is in contrast to that in the case of usual Coulomb interactions, in which intra-orbital couplings are thought to be much larger than interorbital ones. This may be a good starting point to discuss the real lattice systems with two f-electrons per site. 

\section*{Acknowledgments}
 We would like to thank H. Kusunose, H. Kohno and T. Takimoto for valuable comments. This work is supported by a Grant-in-Aid for Scientific Research 
(No.16340103) and the 21st Century COE Program by the Japan Society for the 
Promotion of Science. 
Part of the numerical calculation in this study was carried out at the Yukawa Institute Computer Facility.



\begin{thebibliography}{99}
\bibitem{consider} This expectation comes from the fact that in the pseudospin rotational symmetric case, we can definitely obtain $\tilde{J}_{\pm}=\tilde{J}_{z}$ and the overall features of F's (in magnitudes) are not markedly changed compared with anisotropic cases.
\bibitem{Caut} This is because our $\Gamma_5$ state is generated only by the states with $j_z=\pm5/2$ and $\pm 1/2$ in the j-j coupling scheme, which make our f$^{2}$-$\Gamma_5$ states linear combinations of states with $J=2$ and those with $J=4$. Hund's rule prefers $J=4$ to $J=2$.
\end{thebibliography}
\end{document}